# X-ray Hotspot Flares and Implications for Cosmic Ray Acceleration and magnetic field amplification in Supernova Remnants

Yousaf Butt, *Harvard-Smithsonian Center for Astrophysics, Cambridge, MA 02138, USA*
Troy Porter, *Santa Cruz Institute for Particle Physics, University of California, Santa Cruz, CA 95064, USA*
Boaz Katz & Eli Waxman, *Physics Faculty, Weizmann Institute, Rehovot 76100, Israel*

## ABSTRACT

For more than fifty years, it has been believed that cosmic ray (CR) nuclei are accelerated to high energies in the rapidly expanding shockwaves created by powerful supernova explosions. Yet observational proof of this conjecture is still lacking. Recently, Uchiyama and collaborators reported the detection of small-scale X-ray flares in one such supernova remnant, dubbed "RX J1713-3946" (a.k.a. G347.3-0.5), which also emits very energetic, TeV ($10^{12}$ eV) range, gamma-rays. They contend that the variability of these X-ray "hotspots" implies that the magnetic field in the remnant is about a hundred times larger than normally assumed; and this, they say, means that the detected TeV range photons were produced in energetic nuclear interactions, providing "*a strong argument for acceleration of protons and nuclei to energies of 1 PeV ($10^{15}$ eV) and beyond in young supernova remnants.*" We point out here that the existing multiwavelength data on this object certainly do not support such conclusions. Though intriguing, the small-scale X-ray flares are not the long sought-after "smoking gun" of nucleonic CR acceleration in SNRs.

## INTRODUCTION

Recently, Uchiyama et al. (2007) reported the detection of intriguing small-scale X-ray flares in the supernova remnant (SNR) "RX J1713-3946" (a.k.a. G347.3-0.5), which is also a TeV gamma-ray emitter (Aharonian et al., 2007). They contend that the variability of these X-ray "hotspots" implies that the magnetic field in the remnant is about a hundred times larger than normally assumed; and this, they say, means that the detected

TeV range photons were produced in energetic nuclear interactions, providing "*a strong argument for acceleration of protons and nuclei to energies of 1 PeV ($10^{15}$ eV) and beyond in young supernova remnants.*"

The authors' main argument is that the X-ray variability of the small "hotspot" regions implies a variability of the energetic electron population on the flare time scale (~1yr), which in turn requires a strong magnetic field, ~1mG, to allow the electrons to lose a significant fraction of the energy during this period (Uchiyama et al., 2007). For such a strong magnetic field, the detected X-ray emission would be produced by a small population of electrons, which could not account for the observed TeV flux. The authors therefore conclude that the TeV emission is produced by nuclei, rather than by electrons.

**PROBLEMS WITH THE HADRONIC INTERPRETATION**

However, there are two important caveats to such an interpretation. Firstly, the X-ray flares do not necessarily reflect variations in the electron population at all. These may well be due, for instance, to a change in the intensity of the local magnetic field on time scales much shorter than the electron acceleration or cooling time. In fact, if the cooling time and acceleration efficiency change on short time scales as suggested by Uchiyama et al., this would likely be accompanied (or generated by) a change in the intensity of the local magnetic field. And so in both cases changes in the magnetic field intensity on ~1yr time scales are expected. Secondly, even if the authors' interpretation of a strong, ~1mG, magnetic field is accepted, the inferred field value would apply only to the miniscule X-ray hotspot regions. Since the TeV gamma-ray emission is present over the entire SNR shell (Aharonian et al., 2007), and not limited to the small hotspots, inverse-Compton emission by high energy electrons cannot yet be ruled out as the origin of the bulk of the extended TeV flux (Porter, Moskalenko & Strong 2006; Katz & Waxman, 2008).

Furthermore, the strong large-scale magnetic field advocated by Uchiyama et al. for RX J1713-3946 is incompatible with the very weak extended radio emission of this remnant (Lazendic et al., 2004). Consider the relationship between the predicted radio synchrotron luminosity at 1.4GHz, $\nu L_\nu$ (sync; 1.4GHz), and the TeV gamma ray pion decay luminosity, $\nu L_\nu$ (PP; TeV) (Katz & Waxman, 2008):

$$\nu L_\nu \text{ (sync; 1.4GHz)} \sim \nu L_\nu \text{ (PP; TeV)} (K_{ep}/0.01) (B/100 \mu G)^{3/2} (n/1 \text{ cm}^{-3})^{-1}$$

where $B$ is the magnetic field, n is the target density, and the electron and proton distributions are given by $dN_e/dE = K_{ep} dN_p/dE = K_{ep} C E^{-2}$ with $K_{ep}$ the electron/proton number ratio and $C$ the normalization constant. The observed ratio of fluxes per logarithmic frequency at 1.4GHz and 1 TeV, in RX J1713-3946 is $\nu f_\nu$ (1.4GHz)/$\nu f_\nu$ (TeV) ~0.01. Thus, if the extended magnetic field in the remnant really were as high as proposed (i.e. ~1mG), and if the TeV emission were generated by nucleons, this would lead to radio synchrotron emission that is ~1000 times brighter than has been measured (Figure 1).

Only by invoking far fewer high-energy electrons as compared to protons, with a number ratio, $K_{ep} < 10^{-6}$, could the strong magnetic field model of the authors be brought into line with the radio observations. However, this would be highly unrealistic as such low values of $K_{ep}$ would be roughly four orders of magnitude below that found both in direct measurements of the composition of Galactic CRs, and in observations of other SNRs (Katz & Waxman, 2008; Vink, 2004; Ellison, Berezhko & Baring, 2000, and references therein).

**CONCLUSIONS**

Thus, mG-scale magnetic fields almost certainly cannot be present over the large swaths of the SNR shell where the TeV emission is detected. If small regions of enhanced magnetic field do exist in RX J1713-3946, it is likely that they are embedded in a much weaker extended field (eg. Pohl, Yan and Lazarian, 2005). Consequently, the type of particles generating the extended gamma-ray emission remains unconstrained. In fact, as recent detailed studies indicate, there is good reason to suspect that the bulk of the TeV emission in this SNR is actually leptonic in origin (Porter, Moskalenko & Strong 2006; Katz & Waxman, 2008; Plaga, 2008).

Quite generally, models of this remnant in which the TeV emission is generated by nucleons require unrealistically few accelerated electrons relative to protons with $K_{ep}<10^{-4}$, and predict thermal X-ray Bremsstrahlung emission which is much brighter than the observed X-ray flux (Katz & Waxman, 2008). The latter problem in such models arises independently of the accelerated electron population or magnetic field value adopted.

Hopefully, the debate regarding the nature of the particles responsible for the TeV emission of RX J1713-3946 will be resolved with additional GeV-range gamma-ray data, e.g., from the *Gamma-ray Large Area Space Telescope* (GLAST) satellite which is scheduled to be launched in 2008. These data will cover a critical energy range and will help discriminate between the leptonic or nucleonic origins of the gamma-ray photons (Funk et al., 2007).

## Predicted Radio Flux in Hadronic Model

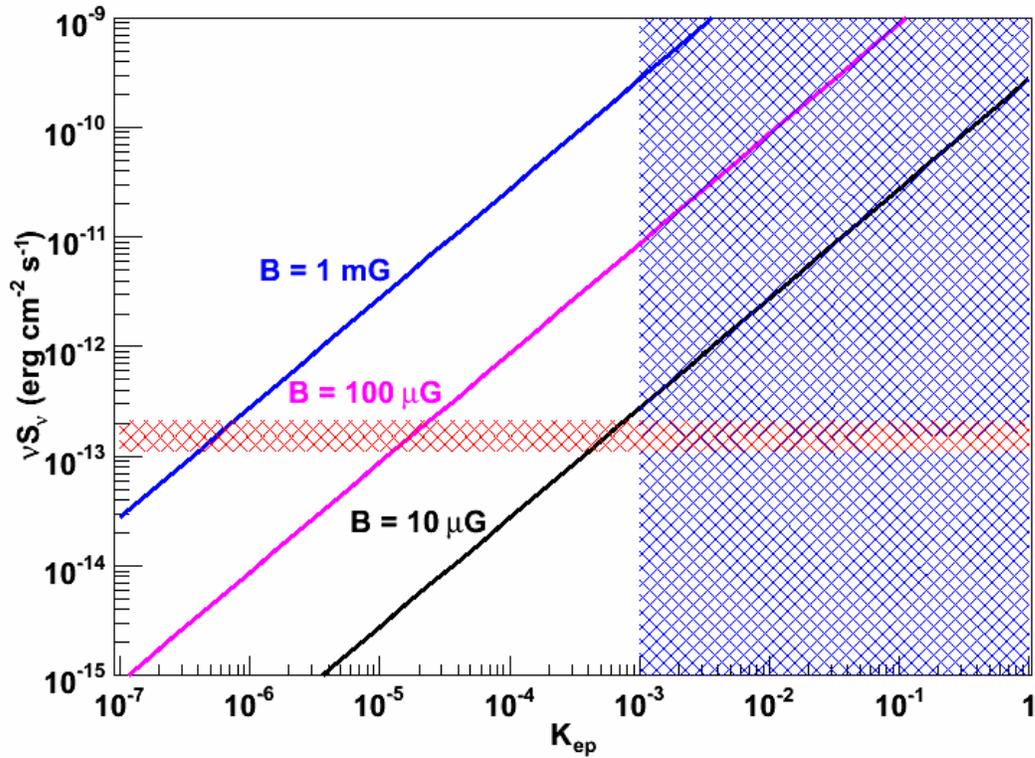

Figure 1: Plot of the model-predicted radio flux at 1.4 GHz as a function of the accelerated electron-to-proton number ratio, $K_{ep}$, calculated for three assumed magnetic fields: $10\mu G$, $100\mu G$, and $1mG$, for the SNR RX J1713-3946. This plot is for a hadronic model, i.e. where the TeV gamma-rays are produced in nuclear interactions, as proposed by Uchiyama et al (2007). The values of $K_{ep}$ are relative to a proton spectrum that reproduces the HESS TeV data (Aharonian et al., 2007), with a target proton density of $0.3$ cm$^{-3}$ (Lazendic et al., 2004). The electron spectrum is of the form $E^{-2}$. The measured value of the radio flux (with ±30% uncertainty) is shown by the red hatched region, and the maximal region of plausible $K_{ep}$ values is shown by the blue hatched region. As can be seen, even for the lowest plausible $K_{ep}$ ratio of $10^{-3}$, the radio flux at 1.4GHz is overpredicted by a factor of ~1000 in the 1mG case proposed by Uchiyama et al. A ratio, $K_{ep}<10^{-6}$, is needed to permit a fit to the detected radio flux if $B=1mG$. But this value of $K_{ep}$ would be a factor of more than ~$10^4$ below what is observed in SNRs, or that found in the CR spectrum, i.e. ~$10^{-2}$ (Katz & Waxman, 2008; Vink, 2004; Ellison, Berezhko & Baring, 2000, and references therein).